\documentclass{ws-procs975x65}
\usepackage{psfrag,curves}

\begin{document}

\title{THEORETICAL GRAVITATIONAL LENSING -- BEYOND THE
WEAK-FIELD SMALL-ANGLE APPROXIMATION}

\author{VOLKER PERLICK}

\address{Physics Department,\\
Lancaster University, \\
Lancaster LA1 4YB, United Kingdom \\
\email{v.perlick@lancaster.ac.uk}}

\begin{abstract}
An overview is given on those theoretical gravitational lensing
results that can be formulated in a spacetime setting, without
assuming that the gravitational fields are weak and that the 
bending angles are small. The first part is devoted to analytical 
methods considering spacetimes in which the equations for light
rays (lightlike geodesics) is completely integrable. This 
includes spherically symmetric static spacetimes, the Kerr
spacetime and plane gravitational waves. The second
part is devoted to qualitative methods which give some information
on lensing properties without actually solving the equation for
lightlike geodesics. This includes Morse theory, methods from
differential topology and bifurcation theory.   
  
\end{abstract}

\bodymatter

\section{Introduction}\label{sec:intro}
For many years, theoretical work on gravitational lensing
was almost exclusively done in an approximation formalism
based on the assumptions that the gravitational fields
are weak and that the bending angles are small. In its
simplest version this formalism gives rise to the 
traditional ``lens equation'', see e.g. Schneider, Ehlers and Falco 
\cite{SchneiderEhlersFalco1992}, Petters, Levine and Wambsganss
\cite{PettersLevineWambsganss2001} or Wambsganss
\cite{Wambsganss1998}. This lens equation has proven a
very powerful tool for investigating gravitational lensing
situations.

In recent years, however, some gravitational lensing situations
have come into the reach of observability for which the
weak-deflection approximation is not applicable.
In particular, the discovery that there is a black hole
at the center of our galaxy, and probably at the center
of most galaxies, is an important motivation for investigating
the deflection of light rays that come close to a black hole.
For such light rays the bending angle may become, in principle, 
arbitrarily large because light rays can make arbitrarily many 
turns around the black hole.
There are also some other, more hypothetical, astrophysical 
objects whose lensing properties may lead to observable effects
beyond the traditional lens equation, such as wormholes or
monopoles. 

If one wants to discuss gravitational lensing without the
weak-field small-angle approximation, one has to use the
formalism of general relativity. In this formalism, light 
rays are mathematically described as lightlike geodesics
of a Lorentzian manifold. The past-oriented lightlike 
geodesics issuing from an observation event $p_O$ make up the 
\emph{past light cone} of $p_O$. All information on what 
can be seen at the celestial sphere of an observer at $p_O$
is coded in the geometry of $p_O$'s past light cone. In this sense,
all lensing features are known if the past light cones are known.
Determining the past light cone of $p_O$ requires solving the
equation for lightlike geodesics through $p_O$. In spacetimes
with sufficiently many symmetries this equation is completely
integrable, so the light cones can be determined analytically. 
In spacetimes where the geodesic equation is not completely
integrable one may study some qualitative features of the
lightlike geodesics without actually solving the geodesic equation,
e.g. with the help of methods from global analysis or 
differential topology. In this sense the theoretical work on gravitational 
lensing in a Lorentzian geometry setting can be naturally 
divided into analytical studies and qualitative studies.

As was already mentioned, the main motivation for investigating
gravitational lensing in a spacetime setting without the weak-field
small-angle approximation comes from observations: There are 
observational situations, accessible with present or near-future
technology, for which this approximation is simply not adequate. 
However, a second and almost equally important motivation
comes from methodology: One gets a much better understanding
of the physics behind gravitational lensing if one develops 
the relevant formalism as far as possible in the general framework
of general relativity and introduces approximations only for
those applications for which they are really necessary.

This article gives an overview on methods of
how to study gravitational lensing in a Lorentzian geometry setting,
without the weak-field small-angle approximation. Chapter 2 reviews
analytical work in spacetimes where the geodesic equation is completely 
integrable. This includes all spherically symmetric static spacetimes 
such as the Schwarzschild spacetime (Section \ref{subsec:spherical}), 
the Kerr spacetime (Section \ref{subsec:kerr}), and plane gravitational 
waves (Section \ref{subsec:wave}). Chapter 3 reviews some mathematical 
methods that give qualitative results on lensing features, e.g. on 
the number of images, without actually solving the geodesic equation. 
This includes Morse theory (Section \ref{subsec:morse}), methods
from differential topology (Section \ref{subsec:topology}) and
bifurcation theory (Section \ref{subsec:bifurcation}). 
Related material can be found in Ref.
\refcite{Perlick2004}.

\section{Analytical Methods}\label{sec:analytical}


\subsection{Spherically Symmetric Static Spacetimes}\label{subsec:spherical}
In a spherically symmetric static spacetime, the geodesic equation
is completely integrable, so the lightlike geodesics can be 
explicitly written in terms of integrals. (In the case of the 
Schwarzschild solution these are elliptic integrals.) This 
allows to write an \emph{exact lens equation} for such spacetimes.

Lensing without weak-field or small-angle approximations in
spherically symmetric static spacetimes was pioneered by Darwin
\cite{Darwin1959, Darwin1961} and by Atkinson  \cite{Atkinson1965}. Whereas 
Darwin's work is restricted to the Schwarz\-schild spacetime 
throughout, Atkinson derives all relevant formulas for an 
unspecified spherically symmetric static spacetime before 
specializing to the Schwarzschild spacetime in Schwarzschild and 
in isotropic coordinates. All important features of Schwarzschild
lensing are clearly explained in both papers. 
However, they do not derive anything like a lens equation.

The first version of a lens equation for spherically symmetric 
static spacetimes that goes beyond the weak-field small-angle 
approximation was brought forward by Virbhadra, Narasimha and 
Chitre \cite{VirbhadraNarasimhaChitre1998} and then, in a modified 
form, by Virbhadra and Ellis \cite{VirbhadraEllis2000}. The 
Virbhadra-Ellis lens equation was originally applied to the 
Schwarzschild spacetime \cite{VirbhadraEllis2000} and later also 
to other spherically symmetric static spacetimes, e.g. to a 
boson star by D\c{a}browski and Schunck \cite{DabrowskiSchunck2000}, 
to a fermion star by Bili{\'c}, Nikoli{\'c} and Viollier 
\cite{BilicNikolicViollier2000}, and to spacetimes with naked 
singularities by Virbhadra and Ellis \cite{VirbhadraEllis2002}.
The Virbhadra-Ellis lens equation might be called an
``almost exact lens equation''. It is approximative insofar as
it restricts to asymptotically flat spacetimes, with observer and 
light source in the asymptotic region and almost aligned (i.e.,
almost in opposite directions from the center of symmetry).
However, the light rays are not restricted to the asymptotic 
region where the gravitational field is weak. They are rather
allowed to enter the central region of the spacetime, on their
way from the source to the observer, and to be arbitrarily 
strongly bent, i.e., to make arbitrarily many turns around the 
center. 

\vspace{-0.8cm}
\begin{figure}
 \centerline{\epsfig{file=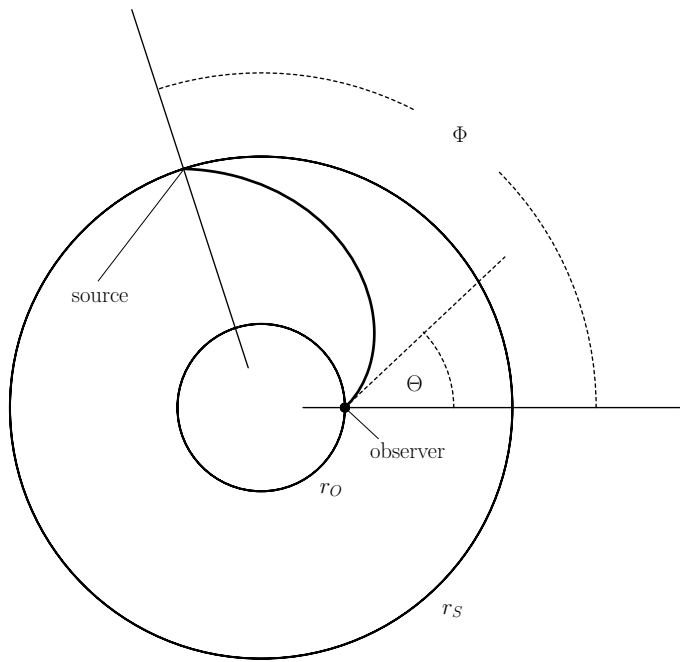,width=0.66\linewidth}} 
\vspace{-0.4cm}
 \caption{Exact lens equation in spherically symmetric static spacetimes.
 The observer is at radius coordinate $r = r_O$,
 the light source at radius coordinate $r = r_S$. The lens 
 equation relates the angle $\Theta$ which gives the image position on 
 the observer's sky to the angle $\Phi$ which gives the source position 
 in the spacetime, see Ref. \protect\refcite{Perlick2004b} for details. 
 \label{fig:sphlens}}
\end{figure}

An exact lens equation for spherically symmetric spacetimes, 
without any restriction on observer or source positions, was
suggested by Perlick \cite{Perlick2004b}. (This can be viewed
as a special case of Fritteli and Newman's exact lens equation 
in arbitrary spacetimes \cite{FrittelliNewman1999},
see Section \ref{subsec:topology}.) To make use of the symmetry, 
one places observer and light source on spheres, rather than on 
planes as in the traditional lens equation, see Fig. \ref{fig:sphlens}. 
The observer is at $r = r_O$ and the light sources are distributed at
$r = r_S$, where $r$ is the radial coordinate in the spherically
symmetric spacetime under consideration. For each past-oriented 
light ray, leaving the observer at an angle $\Theta$ with 
respect to the radial outward direction, one can integrate the
lightlike geodesic equation to get the azimuthal angle 
$\Phi$ swept out by the light ray before it reaches the sphere
$r = r_S$. In this way one gets a lens equation which gives $\Phi$ 
as a function of $\Theta$ and, thereby, relates the source position 
in the spacetime to the image position at the observer's sky. 
Of course, one has to take into account that a light ray with 
$\Phi + 2 \pi$ arrives at the same source position as a light 
ray with $\Phi$. Also, one has to note that light rays may hit the sphere 
$r = r_S$ several times; if this is the case, the lens equation
gives a \emph{multi-valued} map $\Theta \mapsto \Phi$. 

\begin{figure}
 \centerline{\epsfig{file=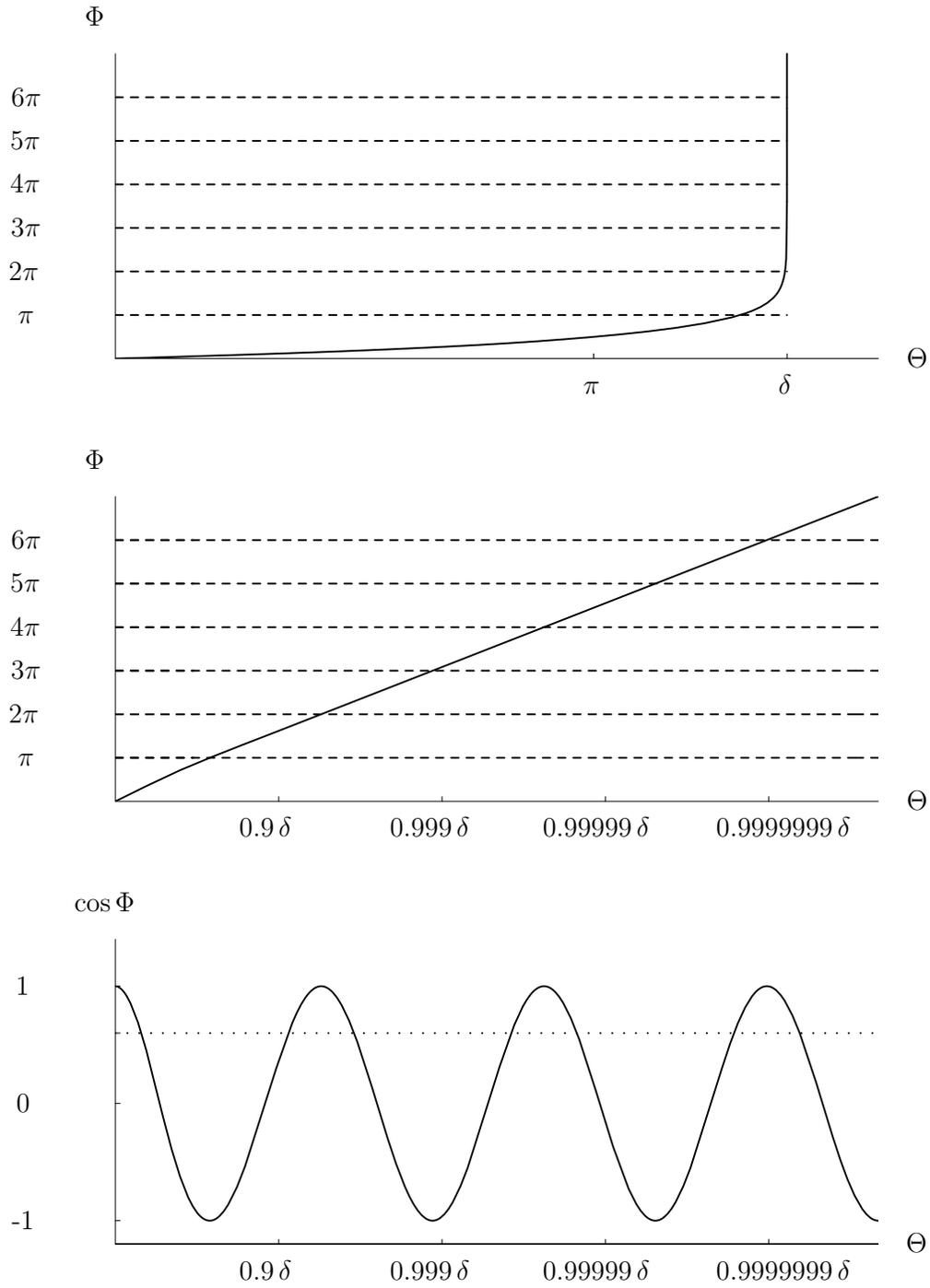,width=1.3\linewidth}} 
 \caption{Plot of $\Phi$ versus $\Theta$ for the Schwarzschild spacetime. 
 \label{fig:schwlens}}
\end{figure}

The lensing features are nicely illustrated if one plots $\Phi$ as a 
function of $\Theta$, as given by the lens equation. This is shown 
for the Schwarzschild spacetime in the top panel of Fig.\ref{fig:schwlens}. 
(Similar pictures for the spacetimes of an Ellis wormhole and of a 
Barriola-Vilenkin monopole can be found in Ref. \refcite{Perlick2004b}.) 
The multiples of $\pi$ are indicated on the $\Phi$ axis; they correspond
to source positions for which the observer sees an Einstein ring. 
If $\Theta$ approaches a certain value $\delta$, the 
angle $\Phi$ becomes infinite, corresponding to a light ray that 
circles around the center forever. This light ray asymptotically 
spirals towards the photon sphere at $r = 3m$. 

A similar behaviour can be observed in any spherically symmetric
static spacetime which is asymptotically flat and has an
unstable photon sphere. (Here the sphere $r= r_L$ is called a 
\emph{photon sphere} if a lightlike geodesic
remains on this sphere if it starts tangentially to it. Thus, a 
photon sphere is filled with circular geodesics. A photon sphere 
is called stable if all lightlike geodesics with initial values
close to those of a circular geodesic stay close to the photon 
sphere; otherwise it is called unstable. The surface $r = 3 m$ 
in the Schwarzschild spacetime is the best known example of an 
unstable photon sphere.) For an observer situated between the 
photon sphere and infinity, the plot of $\Phi$ versus $\Theta$ 
always looks qualitatively as in Fig. \ref{fig:schwlens}. For
a certain value of $\Theta$, the angle $\Phi$ diverges, corresponding 
to a light ray that spirals towards the photon sphere. The divergence
is always logarithmic, as was shown by Bozza \cite{Bozza2002} using
the setting of the Virbhadra-Ellis lens equation. Thus,
by logarithmically rescaling the $\Theta$ axis, the graph of $\Phi$
versus $\Theta$ asymptotically becomes a straight line. For the
Schwarzschild spacetime this is shown in the middle panel of
Fig. \ref{fig:schwlens}. Following Bozza, this straight line is
called the \emph{strong field limit} or the \emph{strong deflection limit}.
(The latter name is actually more appropriate; the bending angle
goes to infinity but the gravitational field, measured in terms 
of the tidal force, need not be particularly strong near a photon 
sphere.) Calculating the strong deflection limit analytically requires
some careful handling of integrals with singularities. The
relevant formulas were worked out  by Bozza \cite{Bozza2002}. 

The fact that $\Phi$ goes to infinity implies that the observer
sees infinitely many images of each light source. This can be 
read from the bottom panel of Fig. \ref{fig:schwlens}: the dotted
line, indicating a particular value of $\mathrm{cos}\, \Phi$, 
intersects the graph of $\mathrm{cos} \, \Phi$ over $\Theta$ 
infinitely many times. For light sources exactly on the axis 
through the observer position ($\,$i.e. $\mathrm{cos} \, \Phi = 
\pm 1 \,$), there are infinitely many Einstein rings. For off-axis 
light sources ($\,$i.e. all other values of $\Phi\,$), there are 
two infinite sequences of isolated images. They correspond to light rays 
that make increasingly many turns around the center, in the positive 
$\varphi$-direction for the first sequence and in the negative 
$\varphi$-direction for the second sequence. Images corresponding 
to light rays that make at least one full turn around the center 
are called \emph{higher-order images}. (The name \emph{relativistic 
images} is also in use, but this name is a bit misleading because 
such images would occur in any theory, relativistic or not, in which 
photons are acted upon by gravity.) Higher-order images have not 
been observed until now, but there is some hope that they might be 
observed sometimes in the future. The most promising candidate for 
such observations is the black hole at the center of our galaxy,
followed by the black hole at the center of M31. The perspectives 
for observing lensed images (secondary and higher order) of stars 
near these black holes were discussed at this conference in a talk 
by Mancini, cf. Refs.  \refcite{BozzaMancini2005, Mancini}. Of course, 
even with the most advanced future technology it will never be 
possible to see all the infinitely many images because there will 
always be some limits on sensitivity and resolution power.

Note that, if $\Phi$ is close to a multiple of $2 \pi$, the light
returns approximately in the same direction from which it came,
after having made one or more turns around the center. This
phenomenon, sometimes called \emph{retrolensing}, was discussed
e.g. by Stuckey \cite{Stuckey1993}, by Holz and Wheeler
\cite{HolzWheeler2002} and by Eiroa and Torres \cite{EiroaTorres2004}. 

\vspace{-0.5cm}
\begin{figure}
\centerline{\epsfig{file=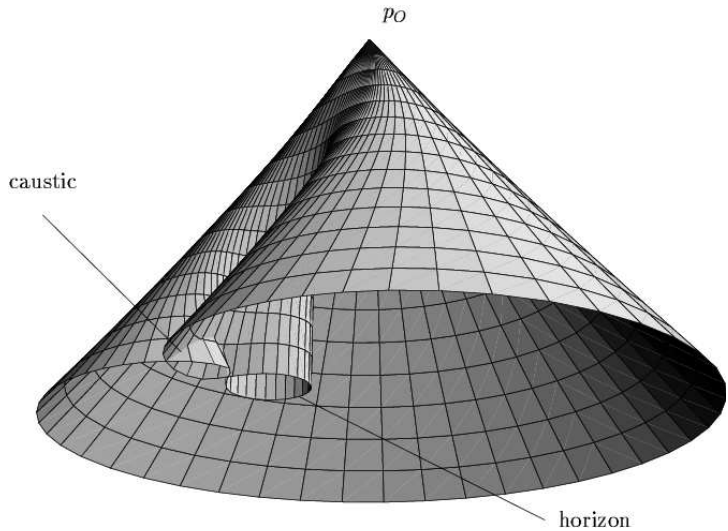,width=0.75\linewidth}} 
 \caption{Past light cone of an event $p_O$ in the Schwarzschild spacetime. 
  This is a $(2+1)-$dimensional spacetime diagram, restricting the spatial 
  coordinates to an equatorial plane. Close to the vertex $p_O$, the light 
  cone looks like the light cone in Minkowski spacetime. Farther away from 
  $p_O$, the light cone wraps around the horizon and forms a caustic which, 
  in this $(2+1)-$dimensional picture appears as a transverse self-intersection 
  of the light cone. Actually, with the missing dimension taken into account,
  at each point of the caustic a circle's worth of light rays intersect. If
  one follows the light cone further down into the past, an infinite sequence
  of caustics is encountered. The second caustic is on the same side of the
  center as $p_O$, the third is opposite, and do on. (For a color version 
  of this picture see Fig. 12 in Ref. \protect\refcite{Perlick2004}.)
 \label{fig:schwcon}}
\end{figure}

If one calculates the strong-deflection limit for observer and light
source in the asymptotic region, one always qualitatively gets the 
same picture as for Schwarzschild, see middle panel of Fig. 
\ref{fig:schwlens}. However, the pitch of the asymptotic straight 
line and its intersection point with the vertical axis are different 
for other spacetimes. Thus, the observation of higher-order 
images would allow to distinguish a Schwarzschild black hole 
from other black holes, see Bozza \cite{Bozza2002} for a detailed 
discussion. For calculations of the strong deflection limit 
in various black hole spacetimes see e.g.  
Refs. \refcite{EiroaRomeroTorres2002, Bhadra2003, 
Eiroa2005a, Eiroa2006} 

The strong deflection limit has also been investigated for wormhole spacetimes
\cite{TejeiroLarranaga2005, NandiZhangZakharov2006}. The qualitative 
lensing features of wormholes are quite similar to black holes. The 
reason is that the spacetime of a wormhole, like that of a black hole, 
is asymptotically flat and has an unstable photon sphere. (The photon sphere 
is at the neck of the wormhole.) As a consequence, the lens map for a wormhole 
is qualitatively similar to that for a black hole if observer and light 
source are in the same asymptotically flat region, compare the 
$\Phi$-versus-$\Theta$ plot for the Schwarzschild spacetime given 
in Fig. \ref{fig:schwlens} with that for an Ellis wormhole 
given in Ref. \refcite{Perlick2004b}.

Iyer and Petters \cite{IyerPetters2006} have recently shown that 
the strong deflection limit in the Schwarzschild spacetime can be 
viewed as the leading order term in an infinite perturbation series. 
A similar perturbation series can be set up with the standard 
weak deflection limit as the leading order term. They are 
able to show that both series together provide an approximation that
deviates by not more than 1 $\%$ from the exact bending angle for
every light ray between the light sphere and infinity.

A completely different approach to analytically evaluating 
the bending angle formula in spherically symmetric static
spacetimes was brought forward by 
Amore et al. \cite{AmoreArceo2006, AmoreArceoFernandez2006, 
AmoreCervantesPaceFernandez2006}. This approach, which makes no 
assumption on either strong or weak bending, converts the formula
for the bending angle into a geometrically convergent series
whose terms can be calculated analytically. Example calculations 
demonstrate that typically only a few terms are needed to get
accurate results, even for cases where the bending angle is
neither particularly small nor particularly large.   

An important motivation for studying gravitational lensing in
spherically symmetric static metrics is in the fact that this
provides a means for testing alternative theories of gravity.
A systematic theoretical framework for investigations of this
kind was brought forward by Keeton and Petters 
\cite{KeetonPetters2005, KeetonPetters2006a, KeetonPetters2006b}
on which Keeton reported at this conference, see Ref. \refcite{Keeton}. 
The basic idea is to write a spherically symmetric static metric as a 
Taylor series with respect to the gravitational radius of the 
central mass. The coefficients of this series are different for
different gravitational theories. The linear term
corresponds to the weak deflection limit; so far all lensing
observations can be satisfactorily explained if only this
term is taken into account. However, as outlined in the quoted 
papers, near-future technology will lead to an accuracy in
observation that requires taking second and higher order terms 
into account. This will give us a new test of Einstein's 
theory and of alternative theories of gravity.

Among alternative theories of gravity, braneworld scenarios are
particularly fashionable. These scenarios provide spherically 
symmetric static black hole solutions for which lensing can 
be studied with the analytical methods discussed above. At this conference 
Whisker (see Refs. \refcite{Whisker2005, Whisker}) and Majumdar 
(see Refs. \refcite{MajumdarMukherjee2005, Majumdar}) reported on 
their recent investigations comparing lensing by braneworld black 
holes to Schwarzschild lensing.

\vspace{-0.3cm}
\begin{figure}
\centerline{\epsfig{file=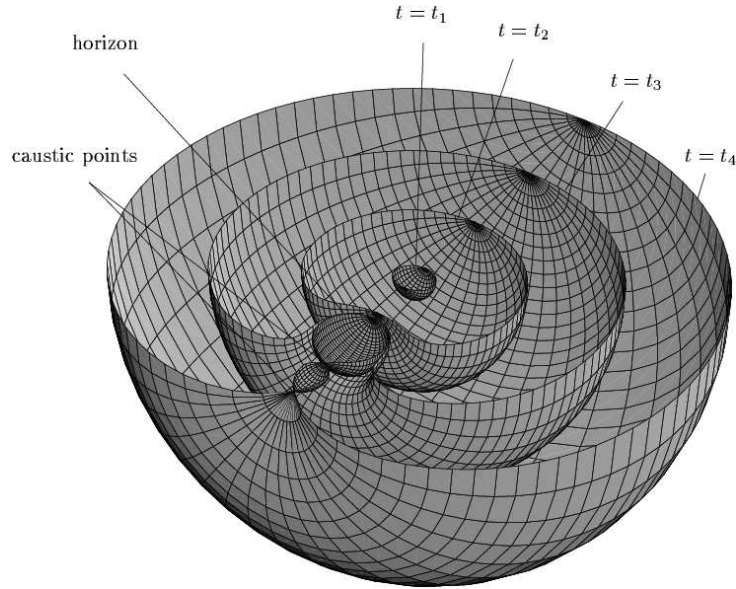,width=0.76\linewidth}} 
 \caption{Wave fronts in the Schwarzschild spacetime. If one intersects the 
 light cone from Fig. \ref{fig:schwcon} with a hypersurface $t = \mathrm{constant}\,$ 
 one gets a wave front. The picture shows such wave fronts, for four different values 
 of $t$, projected to three-space. In contrast to Fig. \ref{fig:schwcon}, all three
 spatial dimemensions can be represented in the picture. If one intersects the 
 light cone close to the vertex, the resulting wave front is a slightly deformed
 sphere. This sphere becomes more and more deformed if the light rays come closer
 to the horizon. After the caustic has formed, the wave fronts are no longer
 topological spheres. (For a color version of this picture see Fig. 13 in Ref.
 \protect\refcite{Perlick2004}.)
 \label{fig:schwfrt}}
\end{figure}


\subsection{Kerr Spacetime}\label{subsec:kerr}

There is strong evidence for a supermassive black 
hole at the center of our galaxy, and some observations
indicate that its spin is non-negligible. A non-vanishing
spin modifies the lensing features of a black hole in a
characteristic way. Some spin-dependent observable effects
were discussed by Zakharov in his talk at this conference,
see Ref. \refcite{Zakharov}.

The appropriate mathematical model for describing the
spacetime around a spinning black hole is the Kerr metric.
Its lightlike (and timelike) geodesics have been studied
in great detail, see e.g. Chandrasekhar \cite{Chandrasekhar1983}
or O'Neill \cite{ONeill1995}. In the Kerr spacetime the geodesic 
equation is completely integrable -- the solutions can be 
written in terms of elliptic integrals -- but the analysis
of the resulting light paths turns out to be rather involved.

The Kerr metric depends on two parameters, the mass $m$ and
the specific angular momentum $a$. In standard Boyer-Lindquist
coordinates, the metric reads
\begin{equation}\label{eq:kerr}
  g =   - \frac{\Delta}{\rho ^2} \, \big( \, dt \, - \, 
  a \, \mathrm{sin} ^2 \vartheta \, d \varphi \big) ^2 \, + \,
  \frac{\mathrm{sin} ^2 \vartheta}{\rho ^2} \, \big(
  (r^2 + a^2) \, d \varphi \, - \, a \, dt \, \big) ^2 \, + \, 
  \frac{\rho ^2}{\Delta} \, dr^2 \,  + \, \rho ^2 \, d \vartheta ^2 \, ,
\end{equation}
where $\rho$ and $\Delta$ are defined by
\begin{equation}\label{eq:rhodelta}
  \rho ^2 = r^2 + a^2 \, {\mathrm{cos}} ^2 \vartheta
  \quad \text{and} \quad 
  \Delta = r^2 - 2mr + a^2  \, .
\end{equation}
For $a=0$ one recovers the Schwarzschild metric. In the following we
restrict to the case $a^2 \le m^2$ which describes a black hole,
and we ignore the case $m^2 < a^2$ which describes a naked 
singularity. 

\vspace{-1.2cm}
\begin{figure}[h]
  \centerline{\epsfig{file=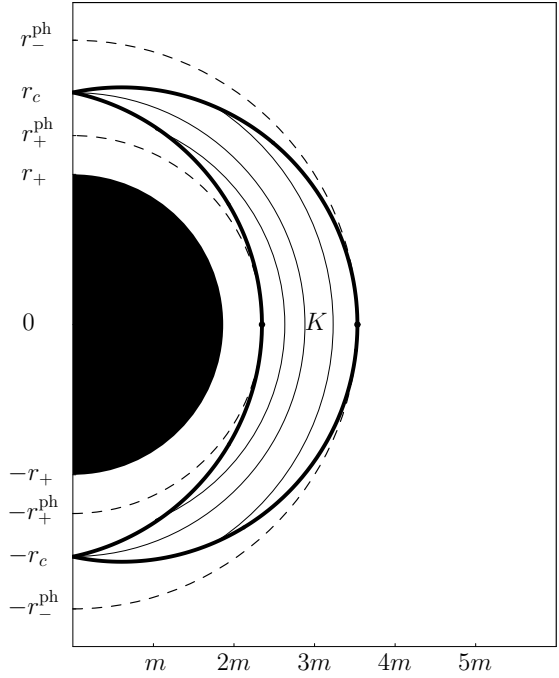,width=0.67\linewidth}} 
\vspace{-1.1cm}
 \caption{The photon region $K$ in the Kerr metric. The picture
  shows a meridional section, i.e., a half plane 
  $\varphi = \,$constant, $t = \,$constant, of the Kerr metric
  for $a = 0.5 \, m$. The outer horizon is at 
  $r_+ $, so the domain of outer communication is the region
  outside of the black ball. The photon region $K$ meets the equatorial 
  plane in two circular lightlike geodesics whose radii 
  $r_+^{\mathrm{ph}}$ and $r_-^{\mathrm{ph}}$ are given by the equation
  $ \big( r_{\pm}^{\mathrm{ph}} \big)^2  - 3 \, m \, r_{\pm}^{\mathrm{ph}}
  + 2 \, a^2 \, = \, \mp \; 
  2 \, a \, \sqrt{ \big( r_{\pm}^{\mathrm{ph}} \big)^2  - 
  2 \, m \,  r_{\pm}^{\mathrm{ph}} \, + \, a^2 \,} \,$, 
  and the axis at radius $r_c$
  given by $r_c^3 \, - \, 3 \, r_c^2 \, m \, + \, r_c \, ( \, a^2  \, + 2 \, q^2 \, )
  \, + \, a^2 \, m \, = \, 0 \, $. 
  \label{fig:KerrK}}
\end{figure}

We denote the region in the Kerr spacetime where
\begin{equation}\label{eq:K}
  \, - \, 2 \, a \, r \, \sqrt{\Delta} \,  {\mathrm{sin}} \, \vartheta  \: \le
  \: 2 \, r \, \Delta - (r-m) \, \rho ^2 \: \le
  \: 2 \, a \, r \, \sqrt{\Delta} \,  {\mathrm{sin}} \, \vartheta 
\end{equation}
by $K$ and we call this the \emph{photon region}, see
Fig. \ref{fig:KerrK}. This region is characterized by the fact that it is filled
with \emph{spherical lightlike geodesics}, i.e., lightlike geodesics that
stay on spheres $r = \,$constant. Along each such geodesic the $\vartheta$
coordinate oscillates between extremal values $- \vartheta _0$ and $\vartheta _0$,
whereas the $\varphi$ and $t$ coordinates increase or decrease linearly with
the affine parameter. In the Schwarzschild limit $a \to 0$ the photon region
$K$ shrinks to the photon sphere $r=3m$.

In the Schwarzschild spacetime light rays that come close to the photon
sphere $r=3m$ can make arbitrarily many turns around the center. Similarly,
in the Kerr spacetime light rays that come arbitrarily close to the photon region
$K$ can make arbitrarily many turns around the center. Thus, one can
define a strong deflection limit in the same sense as for spherically symmetric
static spacetimes. Light rays that start tangential to the equatorial 
plane $\vartheta = \pi /2$ remain in that plane. Their analysis is, 
therefore, not much more difficult than for the spherically symmetric 
static case. One gets infinitely many images on either side of the 
center, but in contrast to the spherically symmetric static
case the positions of the images are not mirror symmetric. This asymmetry
can be interpreted as saying that the light rays are ``dragged'' along 
with the rotation of the black hole. The analytic formulas for the
strong deflection limit in the equatorial plane of the Kerr metric
have been worked out by Bozza \cite{Bozza2003}.

Away from the equatorial plane, the analysis is much more difficult
because then a light ray is not confined to a plane. The strong
deflection limit was calculated by Bozza, de Luca, Scarpetta
and Sereno \cite{BozzadeLucaScarpettaSereno2005} for the case that the 
observer is in the equatorial plane (but the light source need not)
and by Bozza, de Luca and Scarpetta \cite{BozzadeLucaScarpetta2006}
for the general case. These results were presented at this conference
in a talk by de Luca, cf. Ref. \refcite{Luca}. (For another work of 
Kerr lensing without weak-field assumption, largely based on numerical 
studies, see V{\'a}zquez and Esteban \cite{VazquezEsteban2003}.) 
The investigations of the strong deflection limit complements ongoing work
on the weak deflection limit in the Kerr metric which was presented 
by Sereno at this conference, cf. Refs. \refcite{SerenodeLuca2006, Sereno}.

An important aspect of the above-mentioned work on the strong deflection
limit in the Kerr metric is in the fact that it analytically corroborates
knowledge about the caustics in this spacetime which, in earlier work,
has been investigated numerically. If one considers the past light cone 
of an event $p_O$ in the Kerr spacetime, one finds that it forms infinitely
many caustics. By definition, a caustic is a connected subset of the
light cone where neighboring light rays from $p_O$ intersect each other, at least
to first order. At caustic points, the light cone fails to be an
immersed submanifold of the spacetime, forming e.g. cuspidal edges.
In the Kerr spacetime the caustics, if projected to three-dimensional space,
are tubes with astroid cross sections. A picture can be found in
a paper by Blandford and Rauch \cite{RauchBlandford1994} who determined
the caustics numerically. In the Schwarzschild case, $a =0$, the caustics are
just radial lines in three-space. With increasing $a$, they open out into 
tubes with astroid cross-sections, and they start winding around the center. 
In the limit of an extreme Kerr black hole, $a=m$, they spiral asymptotically
towards the horizon. The analysis of the strong deflection limit gives
analytic support to these results which have been found from numerical
studies. However, what is still missing is a good picture that shows how
the caustics are situated inside the light cone. Such pictures can easily 
be produced for the Schwarzschild case in two different ways: First, one can
depict the past light cone of an event in the spacetime, restricting 
oneself to the equatorial plane, see Fig. \ref{fig:schwcon}. Second, 
one can depict in three-space a series of intersections of the 
light cone with hypersurfaces $t=\,$constant (\emph{wavefronts}), see
Fig. \ref{fig:schwfrt}. It would be desirable to have similar
pictures for the Kerr metric. Some attemps in this direction have been 
made, beginning with a 1977 paper by Hanni \cite{Hanni1977}.
At this conference Grave presented computer
visualizations for lensing, including a movie that shows the propagation
of wave fronts in the Kerr spacetime. A picture of wave fronts in
the Kerr spacetime, analogous to the Schwarzschild wave fronts shown
in Fig. \ref{fig:schwfrt}, can be found in Ref. \refcite{Grave}. This picture nicely
shows how the twisting effect, governed by the rotation of the Kerr black hole, 
produces an asymmetry in the wave fronts. However, the resolution of the
picture is too low to show the astroid structure of the caustic. Also, it
is hard to read from this picture the structure of the caustic near the 
horizon. Therefore it seems fair to say that until now the structure of 
the caustics in the Kerr spacetime has not been clearly brought out
in visualizations.

\subsection{Plane Gravitational Waves}
\label{subsec:wave}
Standard textbooks on general relativity usually 
treat gravitational waves as small perturbations 
of Minkowski spacetime. In this weak-field approximation,
it is easy to study the effect of gravitational waves
on light rays. However, the resulting picture is rather
incomplete. Several interesting qualitative features are brought out only if one goes 
beyond the weak-field approximation. This can be done by studying 
plane gravitational waves in terms of exact solutions to Einstein's
field equation. Because of their high symmetry, plane gravitational
wave solutions are  over-idealized in view of applications; however,
they are highly instructive in view of understanding the focusing
properties of gravitational waves.

A \emph{plane gravitational wave} is a spacetime with metric 
\begin{equation}\label{eq:wave}
  g \, = \, -2 \, du \, dv 
   \, - \, 
   \big( \, f(u)\,  (x^2-y^2) \, + \,
   2 \, g(u) \, x 
   \, y \, \big) 
   \,  du^2
   \, + \, dx^2 
   \, + \, dy^2 
\end{equation}
where $f(u)^2+g(u)^2$ is not identically zero. For any choice of $f(u)$ and
$g(u)$, the metric (\ref{eq:wave}) has vanishing Ricci tensor, i.e.,
Einstein's vacuum field equation is satisfied. With the four coordinates
$u,v,x,y$ ranging over all of $\mathbb{R}^4$, the spacetime is 
geodesically complete. The spacelike coordinates $x$ and $y$ 
are transverse to the propagation direction of the gravitational
wave.

\begin{figure}[h]
  \centerline{\epsfig{file=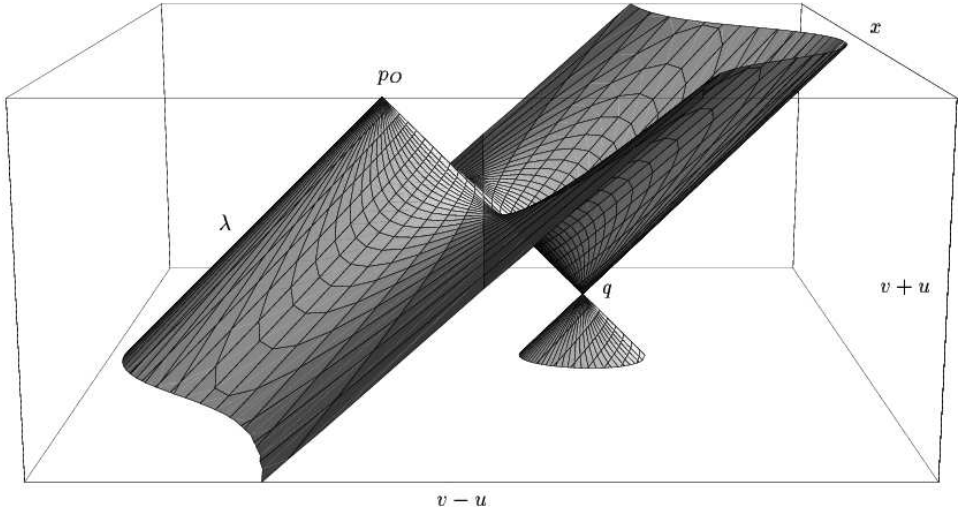,width=1.0\linewidth}} 
 \caption{Past light cone of an event $p_O$ in the spacetime 
  (\ref{eq:wave}) of a plane gravitational wave. With the exception
  of one lightlike geodesic $\lambda$, all the pasr-oriented lightlike geodesics
  from $p_O$ are being refocused in a spacelike curve $q$. The picture was 
  produced with profile functions $f(u)>0$ and $g(u)=0$. Then there 
  is focusing in the $x$-direction and defocusing in the $y$-direction. 
  In the $(2+1)$-dimensional picture, with the $y$-coordinate not shown, 
  the curve $q$ is represented by a point. -- A color version of this picture
  can be found as Fig. 29 in Ref. \refcite{Perlick2004}. For a similar 
  picture, hand-drawn by Roger Penrose, see Ref. \protect\refcite{Penrose1965}.
  \label{fig:wave}}
\end{figure}

For a metric of the form (\ref{eq:wave}), the geodesic equation is 
completely integrable, i.e., the light cone structure can be studied 
excplicitly without approximation. It reveals an interesting
focusing property of gravitational waves that is not captured if 
one restricts to weak fields. This focusing property was discussed
in great detail by Ehrlich and Emch \cite{EhrlichEmch1992, 
EhrlichEmch1993}, cf. Beem, Ehrlich and Easley 
\cite{BeemEhrlichEasley1996}, Chapter~13. What one finds is 
the following. Of all the past-oriented lightlike geodesics issuing
from an event $p_O$, one goes straight to infinity whereas all the
other ones are being refocused in a spacelike curve $q$
which is completely contained in a hypersurface $u = \mathrm{constant}$,
see Fig. \ref{fig:wave}.  The curve $q$ is the first caustic 
of the past light cone of $p_O$. 
This focusing property is universal, i.e., independent of the
profile functions $f(u)$ and $g(u)$. It is closely related 
to the fact that a plane gravitational wave spacetime cannot
be globally hyperbolic. The latter fact was discovered by
Penrose \cite{Penrose1965} who studied the focusing property
for the special case that $f(u)$ and $g(u)$ are different from
zero only in a sufficiently small interval $u_1 < u < u_2$. 
For this special choice of profile functions there is no second
caustic, whereas for other choices of $f(u)$ and $g(u)$ the past-oriented
lightlike geodesics issuing from $p_O$ may be refocused another
time after passing through $q$. It is geometrically evident from
Fig.~\ref{fig:wave} that an inextendible timelike 
curve  $\gamma _S$ which is sufficiently close to $q$ 
intersects the past light cone of $p_O$ exactly twice. 
Thus, an observer at $p_O$ sees 
exactly two images of a light source with worldline $\gamma _S$. 
This example demonstrates that, in contrast to the weak-field theory,
a transparent gravitational lens need not produce an odd number of images, 
even in the case of a geodesically complete spacetime with trivial topology.

\section{Qualitative Methods}\label{sec:qualitative}


\subsection{Morse Theory}\label{subsec:morse}

Morse theory is a body of mathematical theorems that
give some information on the number of solutions to
a variational problem. In applications to lensing,
the variational problem is a version of Fermat's 
principle and the solutions are the light rays 
from the source to the observer. Every such light 
ray corresponds to an image of the source on the
observer's sky. Morse theory can be used to determine
whether the number of images is infinite or finite,
even or odd. Results of this kind are useful because,
if there is an observation which contradicts them,
one knows that one has to search for an additional
image which has gone unnoticed so far. This may 
have happened because an image was too faint or
too close to another image for being detected.

\begin{figure}[h]
   \psfrag{p}{$p_O$}  
    \psfrag{g}{\hspace{-0.2cm} $\gamma _S$}  
   \centerline{\epsfig{file=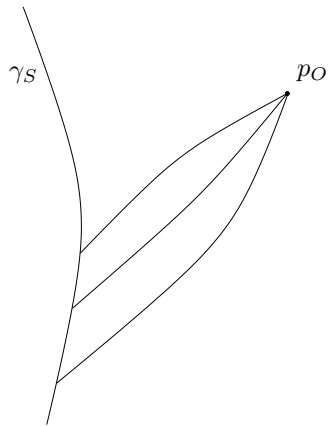,width=0.3\linewidth}} 
 \caption{Kovner's version of Fermat's principle. In an
   arbitrary general-relativistic spacetime, fix an event
   $p_O$ and a timelike curve $\gamma _S$. Consider, as the 
   trial curves, all past-oriented lightlike curves from 
   $p_O$ to $\gamma _S$. The solutions to the 
   variational problem are those trial curves for which
   the arrival time is stationary (i.e., a minimum, a maximum
   or a saddle). Here ``arrival time'' refers to a parametrization
   of $\gamma _S$. This variational problem was brought forward
   by Kovner \protect\cite{Kovner1990}. A proof that the solution
   curves are, indeed, precisely the lightlike geodesics from $p_O$
   to $\gamma _S$ 
   was given by Perlick \protect\cite{Perlick1990b}. A local version
   of this variational principle, restricted to convex normal
   neighborhoods, can be found already in a 1938 paper by
   Temple \protect\cite{Temple1938}.     
  \label{fig:fermat}}
\end{figure}
 
Morse theory can be applied to gravitational 
lensing either in the weak-field small-angle approximation,
or in a spacetime setting without such approximations.
In the following we will be concerned only with
the latter, for the former see Petters \cite{Petters1992}
or the respective chapter in Petters, Levine and Wambsganss
\cite{PettersLevineWambsganss2001}. We will concentrate
on a version of Morse theory that was brought forward 
by Uhlenbeck \cite{Uhlenbeck1975} and applied to 
gravitational lensing first by McKenzie \cite{McKenzie1985}.
This version is restricted to globally hyperbolic
spacetimes throughout. There have been several attempts
to formulate a Morse theory in more general spacetimes,
starting out from Kovner's version \cite{Kovner1990}
of Fermat's principle, see Fig. \ref{fig:fermat}. 
Although this variational principle can be formulated
in an arbitrary spacetime, a fully-fledged Morse 
theory requires additional assumptions which
come close to global hyperbolicity. For the 
most general version available, which applies to
some subsets-with-boundary of a stably causal
spacetime, see Giannoni, Masiello and Piccione
\cite{GiannoniMasielloPiccione1998}.

Following Uhlenbeck \cite{Uhlenbeck1975} we consider a 4-dimensional 
Lorentzian manifold $(M,g)$ that admits a foliation into smooth 
Cauchy surfaces, i.e., a globally hyperbolic spacetime. (The fact 
that the original definition of global hyperbolicity is equivalent 
to the existence of a foliation into \emph{smooth} Cauchy surfaces 
was completely proven only recently by Bernal and 
S{\'a}nchez \cite{BernalSanchez2005}.) Then $M$ can be written 
as a product of a three-dimensional manifold $S$, which serves as the 
prototype for each Cauchy surface, and a time-axis,
\begin{equation}\label{eq:product}
  M= S \times {\mathbb{R}} \, .
\end{equation}
In this spacetime we fix an event $p_O$ and a timelike curve $\gamma _S$,
and we make the following additional assumptions. (We are interested in
past-oriented lightlike geodesics from $p_O$, whereas Uhlenbeck
considered future-oriented lightlike geodesics. Therefore, our
assumptions are the time-reversals of her's.)
\begin{itemize}
\item[(a)]
With respect to the splitting (\ref{eq:product}), the spacetime 
has no particle horizons. (An analytical criterion for this to be
true is the time-reversed version of Uhlenbeck's ``metric growth 
condition''.)
\item[(b)]
The timelike curve $\gamma _S$ is inextendible and it does not meet 
the caustic of the past light cone of $p_O$. (This excludes
situations where the observer sees an extended image, such as
an Einstein ring.)
\item[(c)]
There is a sequence $t_n$ with $t_n \to - \infty$ for $n \to \infty$
such that the projection to $S$ of $\gamma _S (t_n)$ has an accumulation point. 
(Roughly speaking, this condition prohibits that $\gamma _S$ ``goes 
to infinity'' in the past.) 
\end{itemize}
Then the Morse inequalities 
\begin{equation}\label{eq:Morseineq}
  N_k \ge B_k \;  \qquad {\text{for all}} \quad k \in {\mathbb{N}}_0
\end{equation}
  and the Morse relation
\begin{equation}\label{eq:Morserel}
  \sum_{k=0}^{\infty} (-1)^k N_k =
  \sum_{k=0}^{\infty} (-1)^k B_k 
\end{equation}
hold true, where $N_k$ denotes the number of past-pointing lightlike geodesics with
index $k$ from $p_O$ to $\gamma _S$, and $B_k$ denotes the $k$-th Betti number of the loop 
space of $M$. (The index of a geodesic is the number of its conjugate points, counted 
with multiplicity. The loop space of a connected topological space is the space of all 
continuous curves connecting two fixed points. For the definition of Betti numbers
see, e.g., Frankel \cite{Frankel1997}.)

The sum on the right-hand side of (\ref{eq:Morserel}) is, by definition, the
Euler characteristic $\chi$ of the loop space of $M$. Hence, (\ref{eq:Morserel})
can also be written in the form
\begin{equation}\label{eq:N+-}
  N_+ - N_- = \chi \, ,
\end{equation}
where $N_+$ (respectively $N_-$) denotes the number of past-pointing lightlike
geodesics with even (respectively odd) index from $p_O$ to $\gamma _S$.

The Betti numbers of the loop space of $M=S \times {\mathbb{R}}$ are, of
course, determined by the topology of $S$. Three cases are to be distinguished.

\noindent
  {\bf Case A}: $M$ is not simply connected. Then the loop space 
  of $M$ has infinitely many connected components, so $B_0 = \infty$. In this
  situation (\ref{eq:Morseineq}) says that $N_0 = \infty$, i.e., that there 
  are infinitely many past-pointing lightlike geodesics from $p$ to $\gamma$ 
  that are free of conjugate points.

\noindent
  {\bf Case B}: $M$ is simply connected but not contractible 
  to a point. Then for all but finitely many $k \in {\mathbb{N}}_0$ we have
  $B_k > 0$. This was proven in a classical paper by Serre \cite{Serre1951}, 
  cf. McKenzie \cite{McKenzie1985}. In this situation (\ref{eq:Morseineq}) implies 
  $N_k > 0$ for all but finitely many $k$. In other words, for almost 
  every positive integer $k$ we can find a past-pointing lightlike geodesic 
  from $p_O$ to $\gamma _S$ with exactly $k$ conjugate points. Hence, 
  there must be infinitely many past-pointing lightlike geodesics from 
  $p_O$ to $\gamma _S$. 

\noindent
  {\bf Case C}: $M$ is contractible to a point. Then the loop space of 
  $M$ is contractible to a point, i.e., $B_0 =1$ and $B_k = 0$ for $k > 0$. 
  In this case (\ref{eq:N+-}) takes the form $N_+ - N_-
  = 1$ which implies that the total number $N_+ + N_- = 2 N_- + 1$
  of past-pointing lightlike geodesics from $p_O$ to $\gamma _S$ is (infinite
  or) odd.

Case C has relevance for asymptotically simple and empty spacetimes. These
spacetimes describe the gravitational field around a transparent isolated
body, see Hawking and Ellis \cite{HawkingEllis1973} for the formal definition.
It is well known that asymptotically simple and empty spacetimes are globally
hyperbolic, with contractible Cauchy surface, and it is not difficult to verify 
that they have no particle horizons. Also, it follows from the definition of 
an asymptotically simple and empty spacetime that $N_+ + N_-$ must be finite. 
Hence, if $\gamma _S$ satisfies conditions (b) and (c) above, Case C applies and 
says that the total number of images must be odd. This version of the 
``odd number theorem'' makes no assumption on the weakness of the 
gravitational field. (It does not apply to the example of a plane 
gravitational wave, treated in Section \ref{subsec:wave}, because 
the latter is not asymptotically simple.) For further discussions of 
the odd number theorem and Morse theory see McKenzie \cite{McKenzie1985}.

Case B has relevance for black hole spacetimes. In particular, it can be
used to give a general proof that a Kerr-Newman black hole produces 
infinitely many images, with only very mild restrictions on the allowed
motion of the light source. Here the globally hyperbolic spacetime to
be considered is the domain of outer communication of the black hole, 
i.e., the region outside of the outer horizon. Its Cauchy surfaces have
topology $S^2 \times \mathbb{R}$ which is, indeed, simply connected but
not contractible. The details are worked out 
in Ref. \refcite{HassePerlick2005}.

\subsection{Differential Topology}\label{subsec:topology}
The idea of formulating a lens map in an arbitrary general-relativistic 
spacetime, without weak-field or small-angle approximations, is due to 
Frittelli and Newman \cite{FrittelliNewman1999} and was further developed 
by Ehlers, Frittelli and Newman \cite{Ehlers2000, EhlersFrittelliNewman2003}.
In this section we briefly review the construction of their lens map and 
we discuss how methods from differential topology can be applied to it. 

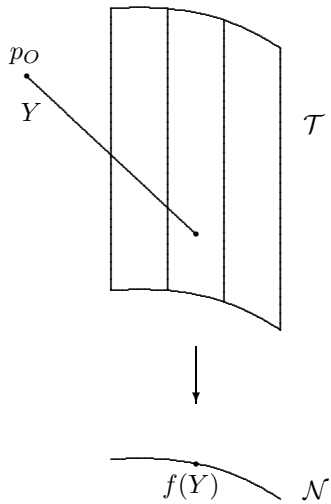
\begin{figure}
\begin{center}
\setlength{\unitlength}{0.75cm}
\begin{picture}(10,10)
	\curve(6,0.5, 5,1, 4,1.2, 3,1.2)
  	\curve(6,3.5, 5,4, 4,4.2, 3,4.2)
	\curve(6,8.5, 5,9, 4,9.2, 3,9.2)
	\curve(6,3.5, 6,8.5)
	\curve(5,4, 5,9)
	\curve(4,4.2, 4,9.2)
	\curve(3,4.2, 3,9.2)
        \curve(1.5,8, 4.5,5.2)        
	\put(1.5,8){\circle*{0.08}}
	\put(4.5,5.2){\circle*{0.08}}
 	\put(4.5,1.13){\circle*{0.08}}     
	\put(6.4,7){${\mathcal{T}}$}
	\put(6.4,0.5){${\mathcal{N}}$}
	\put(1.2,8.3){$p_O$}
        \put(1.4,7.2){$Y$}
	\put(3.9,0.6){$f(Y)$}
	\put(4.5,3.2){\vector(0,-1){1}}
\end{picture}
\end{center}
\vspace{-0.3cm}
\caption{Illustration of the lens map $f$}\label{fig:lensmap} 
\end{figure}

In the standard weak-field small-angle formalism for thin lenses 
the lens map, resulting from the lens equation, is a map from a deflector 
plane to a source plane, see e.g. Schneider, Ehlers and Falco 
\cite{SchneiderEhlersFalco1992}, Petters, Levine and Wambs\-ganss
\cite{PettersLevineWambsganss2001} or Wambsganss \cite{Wambsganss1998}. 
To define something analogous in an arbitrary spacetime, without 
approximations, we fix an observation event $p_O$ and we look for 
analogues of the deflector plane and the source plane. As to the 
deflector plane, there is an obvious candidate, namely the \emph{celestial 
sphere} $\mathcal{S}_O$ at $p_O$. This can be defined as the set of all 
lightlike directions at $p_O$. As to the source plane, however, there is 
no natural candidate. Following Frittelli, Newman and Ehlers, we 
choose a timelike 3-dimensional submanifold $\mathcal{T}$ of the 
spacetime manifold. We assume that $\mathcal{T}$ is of the form
$\mathcal{T} = \mathcal{N} \times \mathbb{R}$ where the $\mathbb{R}-$lines
are timelike and can be interpreted as the worldlines of light sources. 
The two-dimensional manifold $\mathcal{N}$ is the analogue of the
source plane. In this situation the lens map $f : \mathcal{S}_O 
\longrightarrow \mathcal{N}$ is defined by associating each lightlike
direction $Y$ at $p_O$ with the geodesic to which it is tangent,
extending this lightlike geodesic into the past until it meets 
$\mathcal{T}$ and then projecting onto $\mathcal{N}$, see 
Fig. \ref{fig:lensmap}. In general, this prescription does not 
give a well-defined map $f$ since neither existence nor uniqueness of the 
target value is guaranteed. As to existence, there might be some 
past-pointing lightlike geodesics from $p_O$ that never reach $\mathcal{T}$. 
As to uniqueness, one and the same lightlike geodesic might intersect 
$\mathcal{T}$ several times. 

\vspace{-0.5cm}

However, there are some situations of physical interest where the lens 
map is well-defined. An important example is the lens map in asymptotically 
simple and empty spacetimes. This class of spacetimes was mentioned already 
in Section \ref{subsec:morse}, and it was indicated how Morse theory can be 
used to prove that, in such a spacetime, the number of images is odd under
very general conditions. The same result can also be found with the
help of the Frittelli-Newman lens map. The proof, which is worked out in
Ref. \refcite{Perlick2001}, proceeds in two steps. First one considers
``light sources at infinity''; in this case the lens map is a map from
a two-sphere to a two-sphere, and it can be shown that its mapping degree
(Brouwer degree) is equal to $\pm 1$. In the second step it is proven 
that this result implies an odd number of images for very general
light sources in the asymptotically simple and empty spacetime.
Ref. \refcite{Perlick2001} contains a number of other applications
of differential topology to the lens map. 

Clearly, the ``exact lens equation'' for spherically symmetric static
spacetimes, discussed in Section \ref{subsec:spherical}, defines a
lens map which is a special case of the Frittelli-Newman lens map. 
 
\subsection{Bifurcation Theory}\label{subsec:bifurcation}

\begin{figure}
\begin{center}
    \psfrag{C}{$\lambda (s_0 )$} %
    \psfrag{l}{$\lambda$} %
    \psfrag{g}{\hspace{-0.4cm} $\gamma _S$} %
    \psfrag{e}{$\lambda (s_0 - \varepsilon )$} %
    \psfrag{E}{$\lambda _{\varepsilon}$} %
    \psfrag{O}{\hspace{-0.15cm}$\lambda (0)$} %
 \epsfig{file=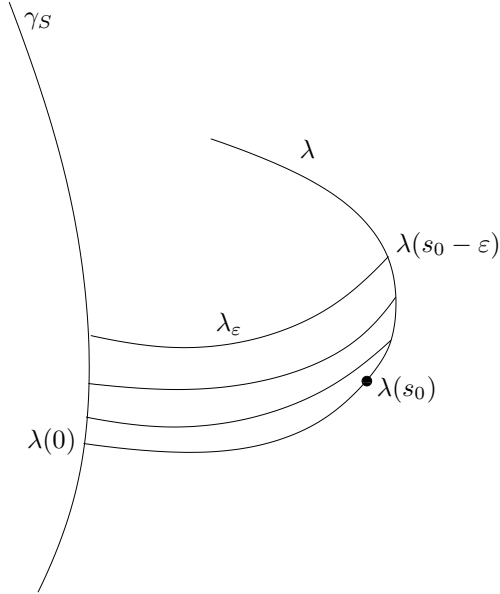,width=0.5\linewidth} 
\end{center}
\vspace{-0.2cm}
 \caption{If $\lambda (s_0)$ is conjugate to $\lambda (0)$,
     events beyond $\lambda (s_0 )$ can be connected to 
     $\gamma _S$ by a second lightlike geodesic.
 \label{fig:bifurcat}}
\end{figure}

Bifurcation theory is concerned with variational problems that
depend on a real parameter. The goal is to characterize the
situation that, in dependence on the parameter, a solution
to the variational problem (i.e., a stationary point of the
variation functional) bifurcates into two or more solutions.
Recently bifurcation theory has been applied to Kovner's 
version of Fermat's principle, recall Fig. \ref{fig:fermat},
by Giamb{\'o}, Giannoni and Piccione \cite{GiamboGiannoniPiccione2004}
and by Javaloyes and Piccione \cite{JavaloyesPiccione2006}. 
To discuss their approach, we fix a timelike curve $\gamma _S$ and 
a past-oriented lightlike geodesic $\lambda$ such that 
$\lambda (0)$ is on $\gamma _S$, see Fig. \ref{fig:bifurcat}. 
Now for every $s<0$ we can consider Fermat's principle with
the given worldline $\gamma _S$ and the observation event 
$p_O = \lambda (s)$. This gives us a one-parameter family of 
variational problems, depending on the parameter $s$. For any 
value of $s$, a section of $\lambda$ is a solution curve of 
the variational problem. With the help of bifurcation theory 
one can characterize those values $s_0$ of $s$ where this 
solution bifurcates into two or more solutions. It is easy 
to see that a bifurcation can occur only at points 
$\lambda (s _0)$ which are conjugate to $\lambda (0)$.
The latter condition means that $\lambda (0)$ is in the caustic
of the past light cone of the event $p_O = \lambda (s _0)$. 
Bifurcation theory gives us some information on conjugate points 
that is difficult to get with other methods. In the following
we present one of the results that is proven in the quoted 
papers.
 
As a preparation, we recall some facts which can be
verified with elementary means. Let $\lambda$ be a past-oriented 
lightlike geodesic and assume that, for some $s_0 < 0$, the 
point $\lambda (s _0)$ is conjugate to $\lambda (0)$, see again
Fig. \ref{fig:bifurcat}. Then
it is an elementary exercise to verify that, for every 
$\varepsilon  > 0$, the geodesic from $\lambda (s _0 - \varepsilon)$ 
to $\lambda (0)$ contains a point conjugate to $\lambda 
(s_0 - \varepsilon)$. Hence a known result (Theorem 2 in 
Ref. \refcite{Perlick1996}) implies that there is a timelike 
curve $\gamma _S$ through $\lambda (0)$ that can be connected 
to $\lambda (s_0 - \varepsilon)$ by a second lightlike geodesic 
$\lambda _{\varepsilon}$.

With the help of bifurcation theory, this statement can be strengthened
in the following way, see Proposition 13 in Ref. \refcite{JavaloyesPiccione2006}. 
For $\varepsilon$ sufficiently small,
\begin{itemize}
\item
the statement is true for \emph{all} timelike curves $\gamma _S$
through $\lambda (0)$;
\item
$\lambda _{\varepsilon}$ depends continuously on $\varepsilon$ (i.e.,
the geodesics $\lambda _{\varepsilon}$ can be parametrized such that
the map $(s, \varepsilon) \mapsto \lambda _{\varepsilon} (s)$ is 
differentiable with respect to $s$ and continuous with respect to 
$\varepsilon$).
\end{itemize}
Thus, an observer at 
$\lambda (s_0 - \varepsilon )$ sees (at least) two images of the light 
source $\gamma _S$, one along the lightlike geodesic $\lambda$ and one 
along the lightlike geodesic $\lambda _{\varepsilon}$. For $\varepsilon 
\to 0$ these two images move continuously towards each other until they 
merge for $\varepsilon = 0$. This property is universal, i.e., it holds 
for all conjugate points.

\vspace{-0.2cm}

\section{Concluding Remarks}\label{sec:conclude}

With an increase in observational accuracy, theoretical 
gravitational lensing beyond the weak-field and small-angle
approximation becomes more and more relevant in view of
applications. At this conference, it was already the 
majority of contributions to the Theoretical Gravitational
Lensing workshop that went beyond these approximations.
Of course, studying lensing in weak gravitational fields
is of remaining value. At this conference this was 
illustrated, in addition to the talks already mentioned,
by Yoo's talk \cite{Yoo} on lensing in a clumpy universe, by
Miranda's talk \cite{Miranda} on lensing by galactic halo 
structures and by Frutos-Alfaro's talk \cite{Frutos} on 
computer programs for visualizing and modeling gravitational 
lenses based on the traditional lens equation. 
Quite generally, weak-field studies have a wide range 
of applications and they complement calculations in 
arbitrarily strong gravitational fields in a useful way. 
From a methodological point of view, it is desirable to 
develop the general formalism in a spacetime setting 
without approximations, as far as possible, and to 
introduce approximations only afterwards for those
applications for which they are necessary.

\bibliographystyle{ws-procs975x65}


\end{document}